\documentclass[10pt]{iopart}
\usepackage{iopams}

\usepackage{graphicx}
\usepackage{color}
\usepackage[english]{babel}
\usepackage[pdfborder={0 0 0}, colorlinks=true, linkcolor=blue, citecolor=blue, urlcolor=cyan]{hyperref}

\def\beq{\begin{equation}}
\def\eeq{\end{equation}}
\def\bea{\begin{eqnarray}}
\def\eea{\end{eqnarray}}

\def\ra{\rangle}
\def\la{\langle}

\def\k{\mathbf{k}}

\def\S{\mathbf{S}}

\newcommand{\sgn}[1] {\mathrm{sgn}\left({#1}\right)}

\newcommand{\abs}[1] {\left|{#1}\right|}
\newcommand{\ii}{{\mathrm{i}}}

\newcommand{\nn}{\nonumber}
\newcommand{\vect}[1] {\mathbf{#1}}

\newcommand{\pp}{{++}}

\newcommand{\Tmat}{\mathcal{T}}
\newcommand{\Uimp}{\mathcal{U}}

\newcommand{\imp}{\mathrm{imp}}

\newcommand{\GammaN}{\Gamma}
\newcommand{\sigmaN}{\sigma}
\newcommand{\etaN}{\eta}

\begin{document}

\title[Details of the $s_\pm \to s_\pp$ transition in 2-band model for Fe-based superconductors]{Details of the disorder-induced transition between $s_\pm$ and $s_\pp$ states in two-band model for Fe-based superconductors}

\author{V.A. Shestakov$^1$, M.M. Korshunov$^{1,2}$, Yu.N. Togushova$^2$, D.V. Efremov$^3$, and O.V. Dolgov$^{4,5}$}
\address{$^1$Kirensky Institute of Physics, Federal Research Center KSC SB RAS, 660036, Krasnoyarsk, Russia}
\address{$^2$Siberian Federal University, 660041, Krasnoyarsk, Russia}
\address{$^3$Leibniz-Institut f\"{u}r Festk\"{o}rper- und Werkstoffforschung, D-01069 Dresden, Germany}
\address{$^4$Max-Planck-Institut f\"{u}r Festk\"{o}rperforschung, D-70569, Stuttgart, Germany}
\address{$^5$P.N. Lebedev Physical Institute RAS, 119991, Moscow, Russia}
\ead{mkor@iph.krasn.ru}

\vspace{10pt}
\begin{indented}
\item[]\date{\today}
\end{indented}

\begin{abstract}
Irradiation of superconductors with different particles is one of many ways to investigate effects of disorder. Here we study the disorder-induced transition between $s_\pm$ and $s_\pp$ states in two-band model for Fe-based superconductors with nonmagnetic impurities. Specifically, the important question of whether the superconducting gaps during the transition change smoothly or steeply? We show that the behavior can be of either type and is controlled by the ratio of intra- and interband impurity scattering and a parameter $\sigmaN$ that represents a scattering strength and changes from zero (Born approximation) to one (unitary limit). For the pure interband scattering potential and $\sigmaN \lesssim 0.11$, the $s_\pm \to s_\pp$ transition is accompanied by the steep behavior of gaps, while for larger values of $\sigmaN$, gaps change smoothly. The steep behavior of the gaps occurs at low temperatures, $T < 0.1 T_{c0}$, otherwise it is smooth. The critical temperature $T_c$ is always a smooth function of the scattering rate in spite of the steep changes in the behavior of the gaps.
\end{abstract}


\vspace{2pc}

\submitto{\SUST}

%
\ioptwocol

\section{Introduction}

Iron-based materials reveal intriguing behavior in a number of physical properties. This includes unconventional superconductivity~\cite{MazinReview,SadovskiiReview2008,HirschfeldKorshunov2011,ReidReview2012,HosonoReview2015,Hirschfeld2016}, transport coefficients and Raman spectra~\cite{StewartReview,b_muschler_09,Canfield2010,KemperKorshunov2011}, magnetic and nematic states~\cite{LumsdenReview,FisherReview2011,Dai2015,Inosov2016}, and electronic band structure~\cite{Ding2008,RichardReview2011,Kordyuk,Zi-RongReview2013,KordyukPseudogapReview}. First one is of a special interest because transition temperature to the superconducting state ($T_c$) is as high as 58~K in bulk materials~\cite{SmFeAsOTc} 
and up to 110~K in a monolayer FeSe~\cite{FeSeTc,FeSeARPES,HeFeSeAnneal,TanFeSeARPES,GeFeSe100K}.

Except for the extreme hole and electron dopings, the Fermi surface of Fe-based materials consists of two or three hole sheets around the $\Gamma=(0,0)$ point and two electron sheets around the $M=(\pi,\pi)$ point of the two-Fe Brillouin zone. Scattering between them with the large wave vector results in the enhanced antiferromagnetic fluctuations, which promote the $s_\pm$ type of the superconducting order parameter that change sign between electron and hole pockets~\cite{MazinReview,HirschfeldKorshunov2011,Korshunov2014eng}. On the other hand, bands near the Fermi level have mixed orbital content and orbital fluctuations enhanced either by vertex corrections or the electron-phonon interaction may lead to the sign-preserving $s_\pp$ state~\cite{Kontani,Bang2009,Onari,Onari2012,Yamakawa2017}. However, most experimental data including observation of a spin-resonance peak in inelastic neutron scattering, the quasiparticle interference in tunneling experiments, and NMR spin-lattice relaxation rate are in favor of the $s_\pm$ scenario~\cite{HirschfeldKorshunov2011,Hirschfeld2016}.

Superconducting states with different symmetries and structures of order parameters act differently being subject to the disorder~\cite{KorshunovUFN2016}. That is, in the single-band $s$-wave superconductor, nonmagnetic impurities do not suppress $T_c$ according to the Anderson's theorem~\cite{Anderson1959}, while the magnetic disorder cause the $T_c$ suppression with the rate following the Abrikosov-Gor'kov theory~\cite{AGeng}. In the unconventional superconductors, suppression of the critical temperature as a function of a parameter $\Gamma$ characterizing impurity scattering may follow a quite complicated law. Several experiments on iron-based materials show that the $T_c$ suppression is much weaker than expected in the framework of the Abrikosov-Gor'kov theory for both nonmagnetic~\cite{Karkin2009,Cheng2010,Li2010,Nakajima2010,Tropeano2010,Kim2014,Prozorov2014} and magnetic disorder~\cite{Cheng2010,Tarantini2010,Tan2011,Grinenko2011,Li2012}. Many theoretical studies revealed the importance of the multiband effects in this matter, see Refs.~\cite{Golubov1997,Ummarino2007,Senga2008,Onari2009,EfremovKorshunov2011,Efremov2013,WangImp2013,KorshunovMagn2014}.
One of the conclusions was that the system having the $s_\pm$ state in the clean case may preserve a finite $T_c$ in the presence of nonmagnetic disorder due to the transition to the $s_\pp$ state. It was obtained both in the strong-coupling $\Tmat$-matrix approximation~\cite{EfremovKorshunov2011} and via a numerical solution of the Bogoliubov-de Gennes equations~\cite{Yao2012,Chen2013}.

Topology of the Fermi surface in Fe-based materials makes it sensible to use a two-band model as a compromise between simplicity and possibility to capture some essential physics. Previously, we have studied the $s_\pm \to s_\pp$ transition in such a model and shown that the transition can take place only in systems with the sizeable effective intraband pairing interaction~\cite{EfremovKorshunov2011}. Physical reason for the transition is quite transparent, namely, if one of the two competing superconducting interactions leads to the state robust against impurity scattering, then although it was subdominating in the clean limit, it should become dominating while the other state is destroyed by the impurity scattering~\cite{KorshunovUFN2016}. Here we focus on the details of the $s_\pm \to s_\pp$ transition. In particular, we are interested in the behavior of the superconducting gaps across the transition. We show that in the case of a weak scattering (including the Born limit) at low temperatures, the gaps behaves steeply, while in all other cases they change smoothly across the transition.

\section{Model}

Hamiltonian of the two-band model can be written in the following form:
\begin{equation}
H = \sum\limits_{\k,\alpha, \sigma} \xi_{\k \alpha} c_{\k \alpha \sigma}^{\dag} c_{\k \alpha \sigma} + \sum\limits_{\mathbf{R}_{i}, \sigma, \alpha, \beta} \Uimp_{\mathbf{R}_{i}}^{\alpha \beta} c_{\mathbf{R}_{i} \alpha \sigma}^{\dag} c_{\mathbf{R}_{i} \beta \sigma} + H_{sc},
\label{eq.H}
\end{equation}
where $c_{\k \alpha \sigma}$ is the annihilation operator of the electron with a momentum $\k$, spin $\sigma$, and a band index $\alpha$ that equals to $a$ (first band) or $b$ (second band), $\xi_{\k \alpha}$ is the electron dispersion that, for simplicity, we treat as a linearized one near the Fermi level, $\xi_{\k \alpha} = \vect{v}_{F \alpha} (\k - \k_{F \alpha})$, with $\vect{v}_{F\alpha}$ and $\k_{F \alpha}$ being the Fermi velocity and the Fermi momentum of the band $\alpha$, respectively. Presence of disorder is described by the nonmagnetic impurity scattering potential $\Uimp$ at sites $\mathbf{R}_{i}$.

Superconductivity occurs in our system due to the interaction $H_{sc}$ that in general can have different forms for different pairing mechanisms. Hereafter we assume that the problem of finding the effective dynamical superconducting interaction is already solved and both coupling constants and the bosonic spectral function are obtained. Latter describes the effective electron-electron interaction via an intermediate boson. In the case of local Coulomb (Hubbard) interaction~\cite{Castallani1978,Oles1983}, intermediate excitations are spin or charge fluctuations~\cite{BerkSchrieffer}, while in the case of electron-phonon interaction, those are phonons. Nature of the effective dynamical interaction is not important for the following analysis; rather important is that the corresponding bosonic spectral function is peaked at some small frequency and drops down with increasing frequency.

\section{Method}

Here we employ the Eliashberg approach for multiband superconductors~\cite{allen}. Dyson equation, $\hat{\mathbf{G}}(\k,\omega_n) = \left[\hat{\mathbf{G}}_0^{-1}(\k,\omega_n) - \hat{\mathbf{\Sigma}}(\k,\omega_n)\right]^{-1}$, establish connection between the full Green's function $\hat{\mathbf{G}}(\k,\omega_n)$, the `bare' Green's function (without interelectron interactions and impurities),
\beq
 \hat{G}_0^{\alpha \beta}(\k,\omega_n) = \left[ \ii \omega_n \hat{\tau}_{0} \otimes \hat{\sigma}_{0} - \xi_{\k \alpha} \hat{\tau}_{3} \otimes \hat{\sigma}_{0} \right]^{-1} \delta_{\alpha \beta}
\eeq
and the self-energy matrix $\hat{\mathbf{\Sigma}}(\k,\omega_n)$. Green's function of the quasiparticle with momentum $\k$ and Matsubara frequency $\omega_n = (2 n + 1) \pi T$ is a matrix in the band space (indicated by bold face) and Nambu space (indicated by hat). Latter denoted by Pauli matrices $\hat{\tau}_{i}$.

Further we assume that the self-energy does not depends on the wave vector $\k$ but keep dependence on the frequency and band indices,
\beq
 \hat{\mathbf{\Sigma}}(\omega_n) = \sum_{i=0}^{3} \Sigma_{(i) \alpha \beta}(\omega_n) \hat{\tau}_i.
\eeq
In this case, the problem can be simplified by averaging over $\k$. Thus, all equations will be written in terms of quasiclassical $\xi$-integrated Green's functions represented by a $4 \times 4$ matrices in Nambu and band spaces,
\beq
\hat{\mathbf{g}}(\omega_n) = \int d \xi \hat{\mathbf{G}}(\k, \omega_n) =
\left(
\begin{array}{cc}
\hat{g}_{an} & 0 \\
0 & \hat{g}_{bn}
\end{array}
\right),
\label{eq.g}
\eeq
where
\beq
 \hat{g}_{\alpha n} = g_{0\alpha n} \hat{\tau}_{0} + g_{2\alpha n} \hat{\tau}_{2}.
 \label{eq.g.alpha}
\eeq
Here, $g_{0\alpha n}$ and $g_{2\alpha n}$ are the normal and anomalous (Gor'kov) $\xi$-integrated Green's functions in the Nambu representation,
\beq
g_{0\alpha n} = -\frac{\ii \pi N_{\alpha} \tilde{\omega}_{\alpha n}}{\sqrt{\tilde{\omega}_{\alpha n}^{2}+\tilde{\phi}_{\alpha n}^{2}}}, \;\;\; g_{2\alpha n} = -\frac{\pi N_{\alpha} \tilde{\phi}_{\alpha n}}{\sqrt{\tilde{\omega}_{\alpha n}^{2}+\tilde{\phi}_{\alpha n}^{2}}}.
\label{eq.g02}
\eeq
They depend on the density of states per spin at the Fermi level of the corresponding band ($N_{a,b}$), and on the renormalized (by the self-energy) order parameter $\tilde{\phi}_{\alpha n}$ and frequency $\tilde{\omega}_{\alpha n}$,
\begin{eqnarray}
\ii \tilde\omega_{\alpha n} &=& \ii \omega_n - \Sigma_{0\alpha}(\omega_n) - \Sigma_{0\alpha}^{\imp}(\omega_n), \label{eq.omega.tilde} \\
\tilde\phi_{\alpha n} &=& \Sigma_{2\alpha}(\omega_n) + \Sigma_{2\alpha}^{\imp}(\omega_n). \label{eq.phi.tilde}
\end{eqnarray}
Often, it is convenient to introduce the renormalization factor $Z_{\alpha n} = \tilde{\omega}_{\alpha n} / \omega_n$ that enters the gap function $\Delta_{\alpha n} = \tilde{\phi}_{\alpha n} / Z_{\alpha n}$. It is the gap function that generates peculiarities in the density of states.

A part of the self-energy due to spin fluctuations or any other retarded interaction (electron-phonon, retarded Coulomb interaction) can be written in the following form:
\bea
\Sigma_{0\alpha}(\omega_n) &=& T \sum\limits_{\omega_n',\beta} \lambda^{Z}_{\alpha\beta}(n-n') \frac{g_{0\beta n'}}{N_\beta}, \label{eq:SigmaSF0} \\
\Sigma_{2\alpha}(\omega_n) &=& -T \sum\limits_{\omega_n',\beta} \lambda^{\phi}_{\alpha\beta}(n-n') \frac{g_{2\beta n'}}{N_\beta},
\label{eq:SigmaSF2}
\eea
Coupling functions,
\beq
 \lambda^{\phi,Z}_{\alpha\beta}(n-n') = 2 \lambda^{\phi,Z}_{\alpha\beta} \int^{\infty}_{0} d\Omega \frac{\Omega B(\Omega)}{(\omega_n-\omega_{n'})^{2} + \Omega^{2}}, \nn
\eeq
depend on coupling constants $\lambda^{\phi,Z}_{\alpha \beta}$, which include density of states $N_{\beta}$ in themselves, and on the normalized bosonic spectral function $B(\Omega)$~\cite{ParkerKorshunov2008,Popovich2010,Charnukha2011}. The matrix elements $\lambda^\phi_{\alpha \beta}$ can be positive (attractive) as well as negative (repulsive) due to the interplay between spin fluctuations and electron-phonon coupling~\cite{BerkSchrieffer,ParkerKorshunov2008}, while the matrix elements $\lambda^Z_{\alpha \beta}$ are always positive. For the simplicity we set $\lambda^Z_{\alpha \beta} = |\lambda^\phi_{\alpha \beta}| \equiv |\lambda_{\alpha \beta}|$ and neglect possible anisotropy in each order parameter $\tilde\phi_{\alpha n}$. 

We use a noncrossing, or $\Tmat$-matrix, approximation to calculate the impurity self-energy $\hat{\mathbf{\Sigma}}^{\imp}$:
\beq
 \hat{\mathbf{\Sigma}}^{\imp}(\omega_n) = n_{\imp} \hat{\mathbf{U}} + \hat{\mathbf{U}} \hat{\mathbf{g}}(\omega_n) \hat{\mathbf{\Sigma}}^{\imp}(\omega_n),
 \label{eq.tmatrix}
\eeq
where $n_{\imp}$ is the concentration of impurities and $\hat{\mathbf{U}}$ is the matrix of the impurity potential. Latter is equal to $\hat{\mathbf{U}} = \mathbf{U} \otimes \hat\tau_3$, where $(\mathbf{U})_{\alpha \beta} = \Uimp_{\mathbf{R}_{i}}^{\alpha \beta}$. Without loss of generality we set $\mathbf{R}_{i} = 0$ for the single impurity problem studied here. For simplicity intraband and interband parts of the impurity potential are set equal to $v$ and $u$, respectively, such that $(\mathbf{U})_{\alpha \beta} = (v-u) \delta_{\alpha \beta} + u$. Relation between the two will be controlled by the parameter $\etaN$:
\beq
 v = u \etaN.
\eeq

There are two important limiting cases: Born limit (weak scattering) with $\pi u N_{a,b} \ll 1$ and the opposite case of a very strong impurity scattering (unitary limit) with $\pi u N_{a,b} \gg 1$. With this in mind, it is convenient to introduce the generalized cross-section parameter
\beq
\sigmaN = \frac{\pi^2 N_a N_b u^2}{1 + \pi^2 N_a N_b u^2} \to \left\{
\begin{array}{l}
 0, \mathrm{Born} \\
 1, \mathrm{unitary}
\end{array}
\right.
\eeq
and the impurity scattering rate
\beq
\GammaN_{a,b} = \frac{2 n_{\imp} \sigmaN}{\pi N_{a,b}} \to \left\{
\begin{array}{l}
 2 n_{\imp}\pi N_{b,a} u^2, \mathrm{Born} \\
 2 n_{\imp}/\left( \pi N_{a,b} \right), \mathrm{unitary}
\end{array}
\right.
\eeq

The procedure of further calculations is the following: i) solve equation~(\ref{eq.tmatrix}), ii) calculate renormalizations of frequency~(\ref{eq.omega.tilde}) and order parameter~(\ref{eq.phi.tilde}) self-consistently, iii) use them to obtain Green's functions~(\ref{eq.g02}) and, consequently,~(\ref{eq.g}).

To determine $T_c$, we solve linearized equations for the order parameter and the frequency,
\bea
 \sum\limits_{\omega_{n'}, \beta} \Bigl[ &\delta_{nn'}&\delta_{\alpha \beta} - \delta_{nn'}\tilde\Gamma_{\alpha \beta} \frac{\sgn{\omega_{n'}}}{\abs{\tilde\omega_{\beta n'}}} \Bigr. \nn\\
 &-& \Bigl. \pi T_c \sum\limits_{\omega_{n'}, \beta}\lambda_{\alpha \beta}(n-n') \frac{\sgn{\omega_{n'}}}{\abs{\tilde\omega_{\beta n'}}} \Bigr] \tilde\phi_{\beta n'} = 0, \label{eq.Lin1} \\
 \tilde\omega_{\alpha n} &=& \omega_n + \sum\limits_{\beta}\tilde\Gamma_{\alpha \beta} \sgn{\omega_n} \nn\\
 &+& \pi T_c \sum\limits_{\omega_{n'}, \beta}\abs{\lambda_{\alpha \beta}(n-n')}\sgn{\omega_{n'}}, \label{eq.Lin2}
\eea
Here $\tilde\Gamma_{\alpha \beta}$ are the components of the impurity scattering rate matrix~\cite{EfremovKorshunov2011},
\beq
 \tilde\Gamma_{a b(b a)} = \frac{\Gamma_{a(b)}(1-\sigma)}{\sigma(1-\sigma)\eta^2 N^2/(N_a N_b)+(\sigma\eta^2-1)^2},
 \label{eq.Gamma_ab}
\eeq
where $N = N_a + N_b$ is the total density of states in the normal phase. Note that the diagonal terms, $\tilde\Gamma_{aa}$ and $\tilde\Gamma_{bb}$, are absent in equations~(\ref{eq.Lin1}) and~(\ref{eq.Lin2}). Equation~(\ref{eq.Lin1}) can be written in the matrix form as $\hat\mathbf{K}\tilde\mathbf{\phi} = 0$, where $\hat\mathbf{K}$ and $\tilde\mathbf{\phi}$ are a matrix and a vector, respectively, in the combined band and Matsubara frequency spaces. 
By varying $T_c$ as a parameter, we determine its value as a point where the sign of $\mathrm{det}|\hat\mathbf{K}|$ changes.

\section{Results}

\begin{figure}
 \centering
 \includegraphics[width=0.5\textwidth]{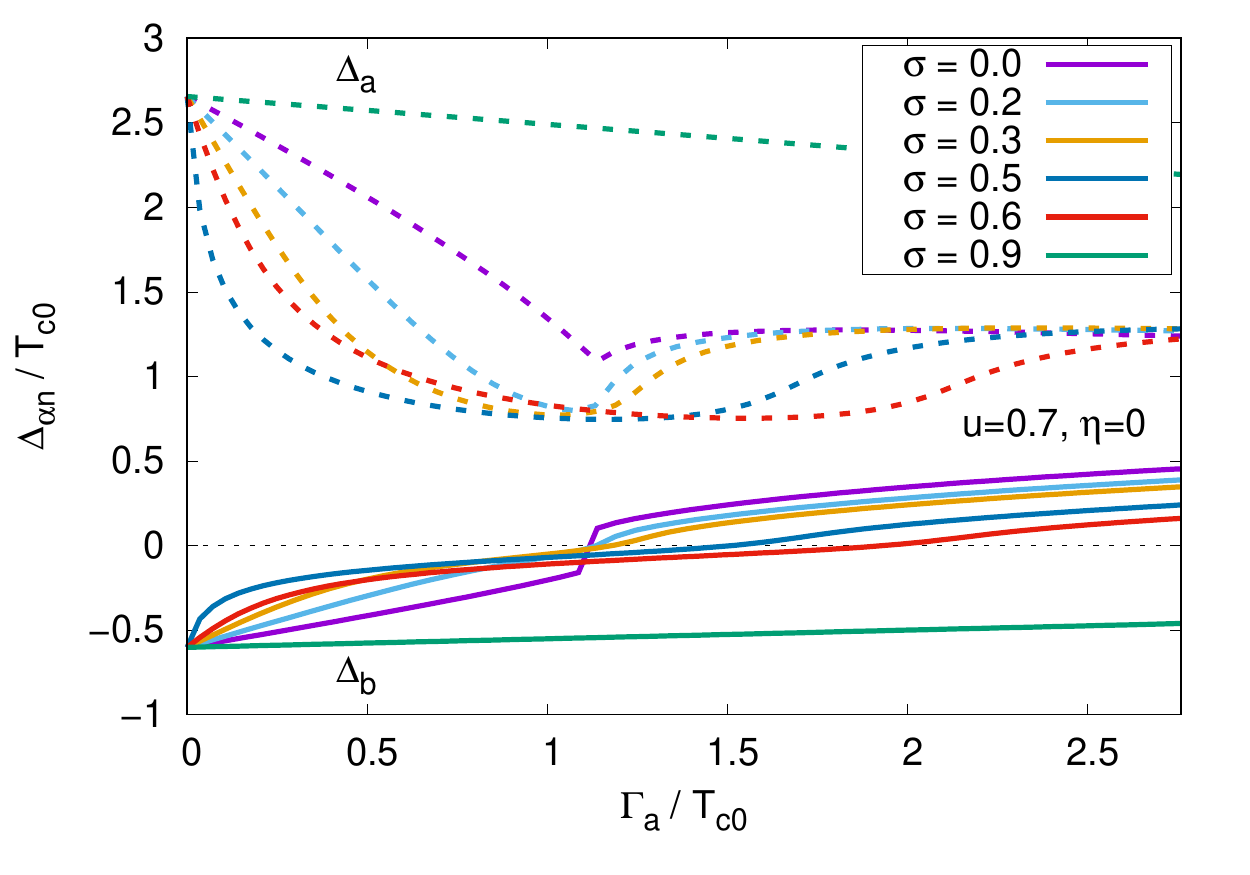}
 \includegraphics[width=0.5\textwidth]{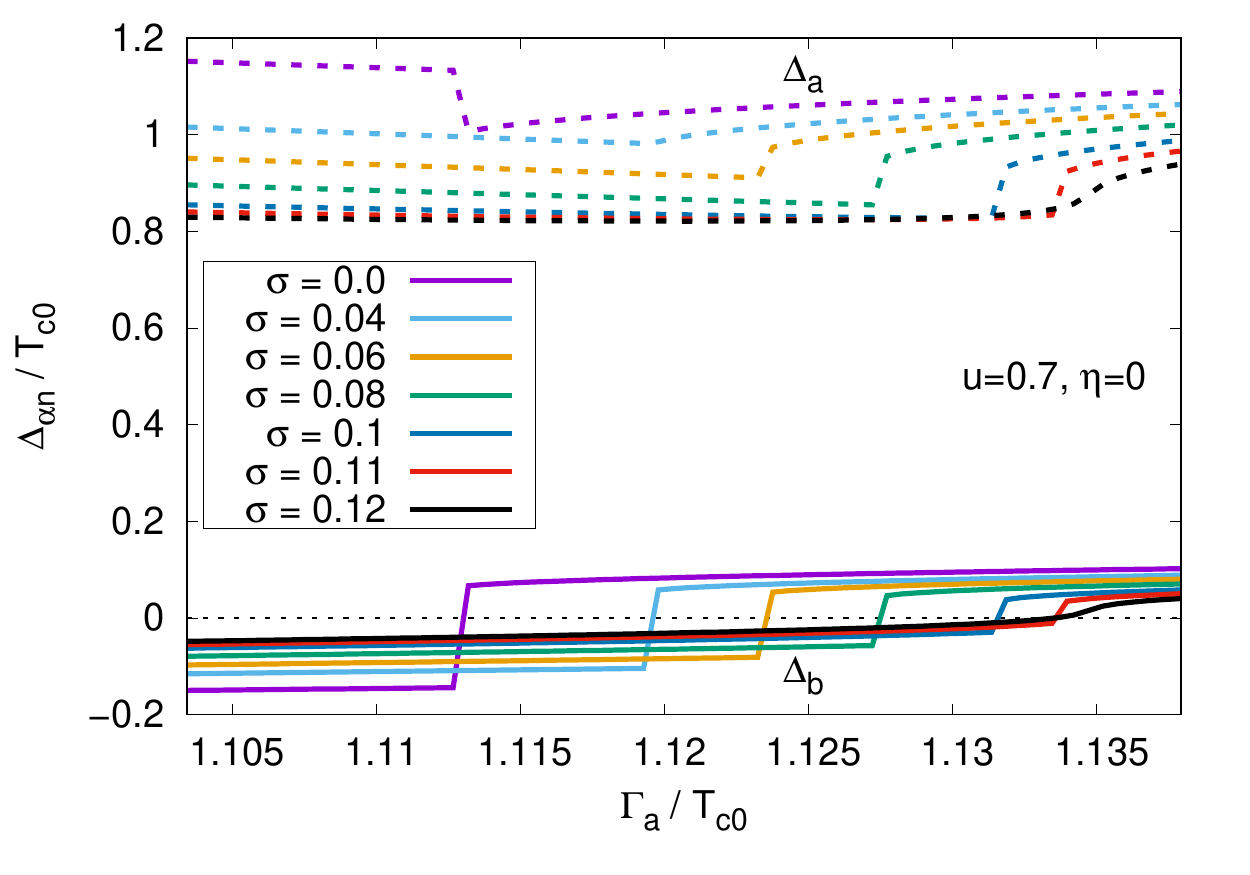}
 \caption{Matsubara gaps $\Delta_{\alpha n=1}$ dependence on the nonmagnetic impurity scattering rate $\GammaN_a$ for the $s_\pm$ state with $\etaN = 0$ and $T = 0.03 T_{c0}$. In the upper panel, we show a wide range of $\sigma$'s, while the data in lower panel demonstrate a jump in gaps at the point of the $s_\pm \to s_\pp$ transition for small values of $\sigmaN$. Note the smooth behavior of gaps for $\sigmaN > 0.11$.
 \label{fig:spmDeltaEta0}}
\end{figure}
\begin{figure}
 \centering
 \includegraphics[width=0.5\textwidth]{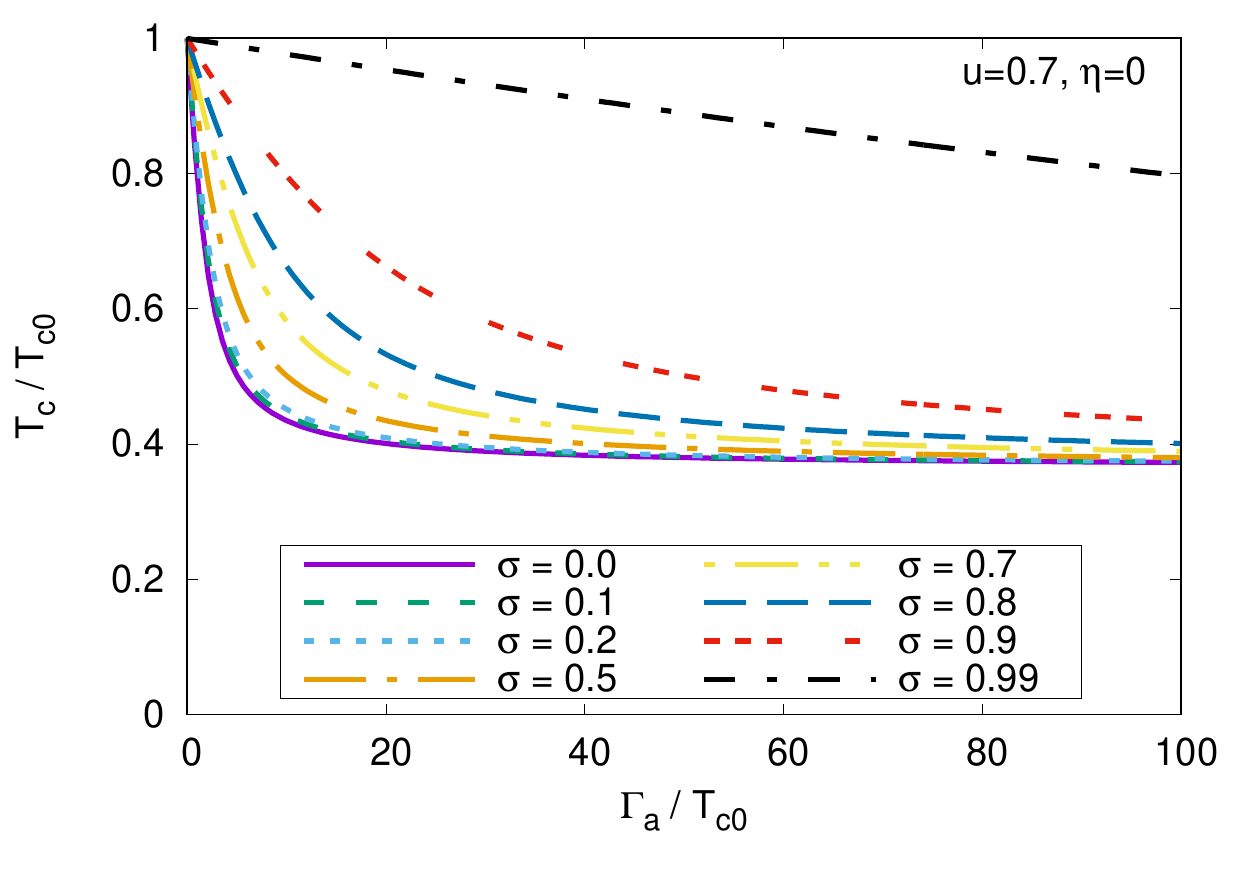}
 \includegraphics[width=0.5\textwidth]{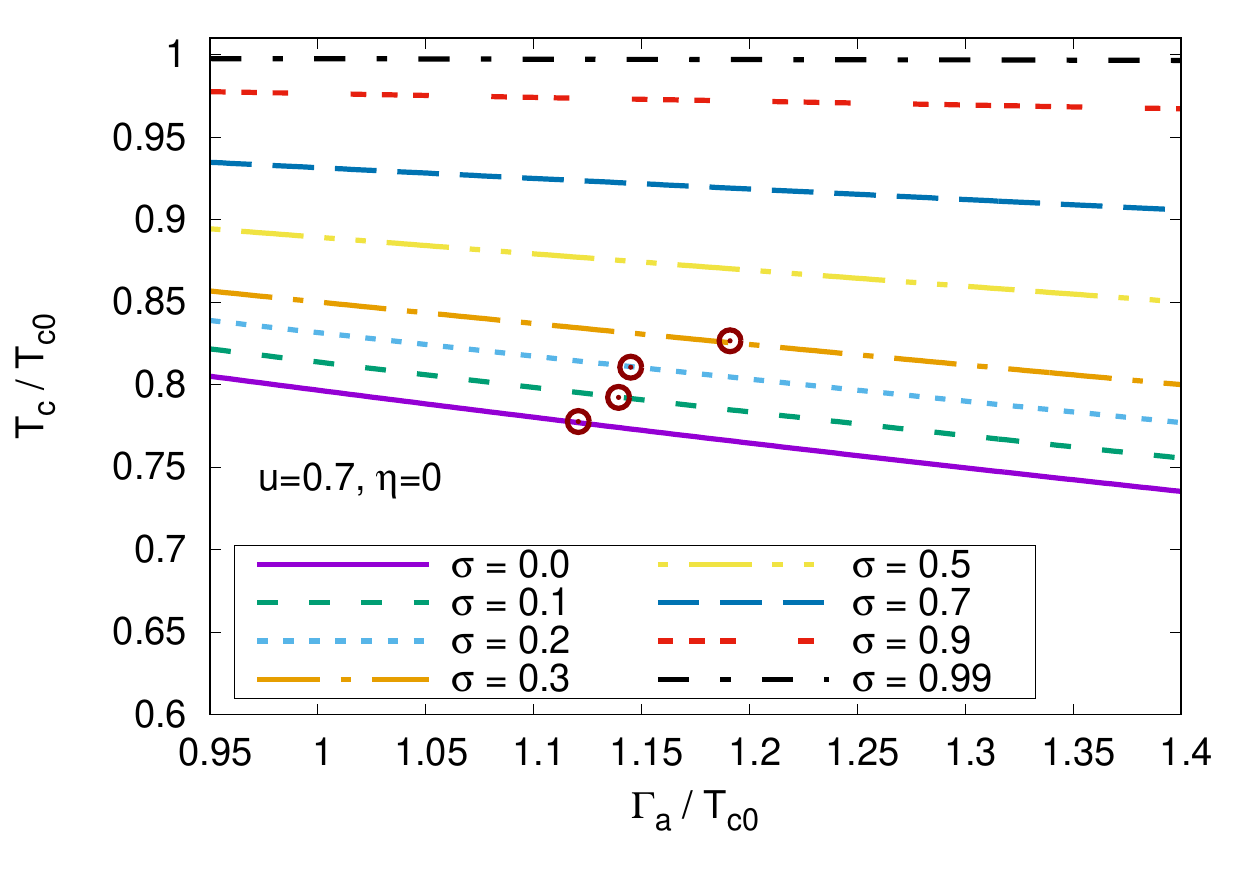}
 \caption{$T_c$ vs. scattering rate $\GammaN_a$ for the $s_\pm$ state with $\etaN = 0$. Results for a wide range of $\GammaN_a$'s is shown in the upper panel. In the lower panel, we show a narrow range of $\GammaN_a$'s where the $s_\pm \to s_\pp$ transition with jumps in gaps takes place. $T_c$ reveals smooth behavior at points of transition, which are marked by circles.
 \label{fig:spmTcEta0}}
\end{figure}
\begin{figure}
 \centering
 \includegraphics[width=0.5\textwidth]{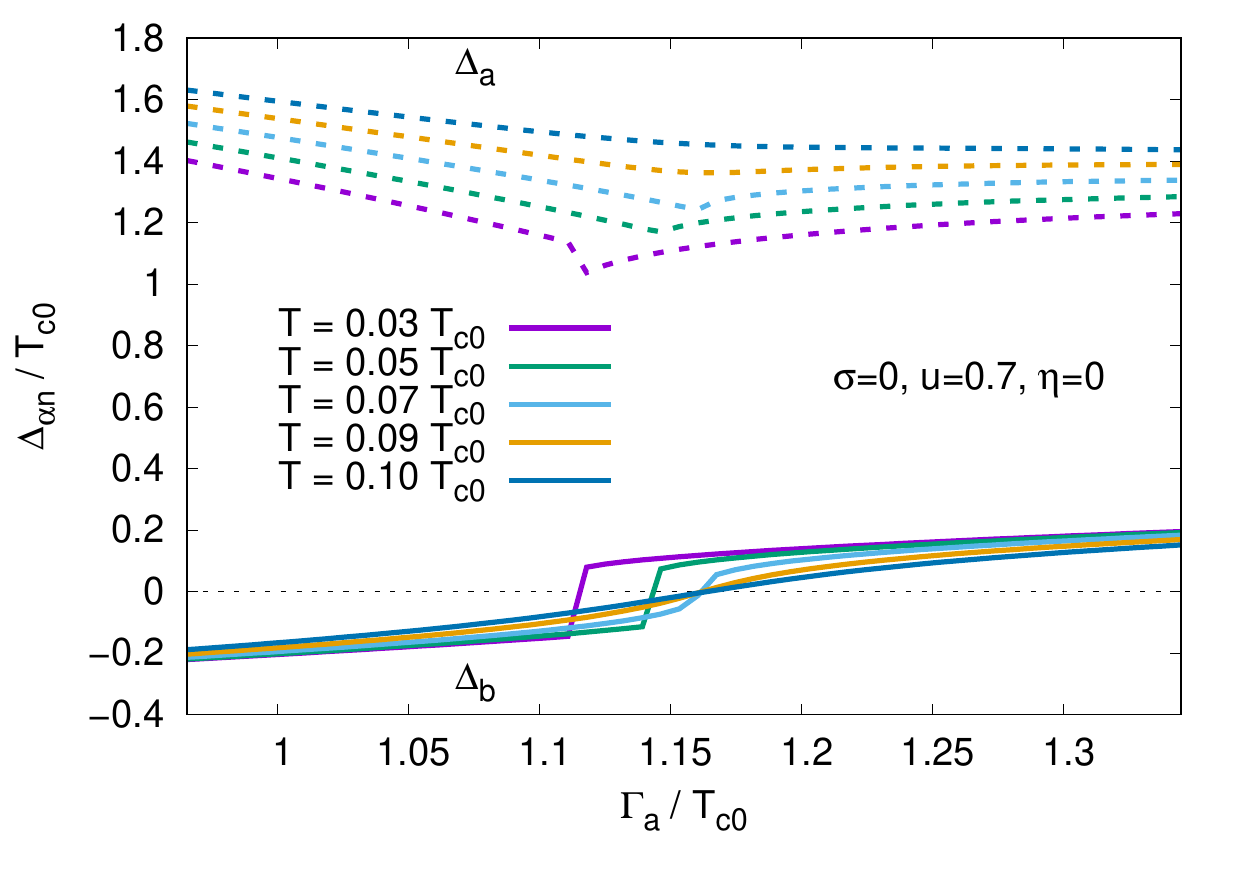}
 \caption{Temperature evolution of Matsubara gaps $\Delta_{\alpha n=1}$ for a range of $\GammaN_a$ with $\sigmaN = 0$ and $\etaN = 0$. Note the steep behavior of smaller gaps at low temperatures and restoration of the smooth behavior for $T \gtrsim 0.09 T_{c0}$.
 \label{fig:spmDeltaT}}
\end{figure}
\begin{figure}
 \centering
 \includegraphics[width=0.48\textwidth]{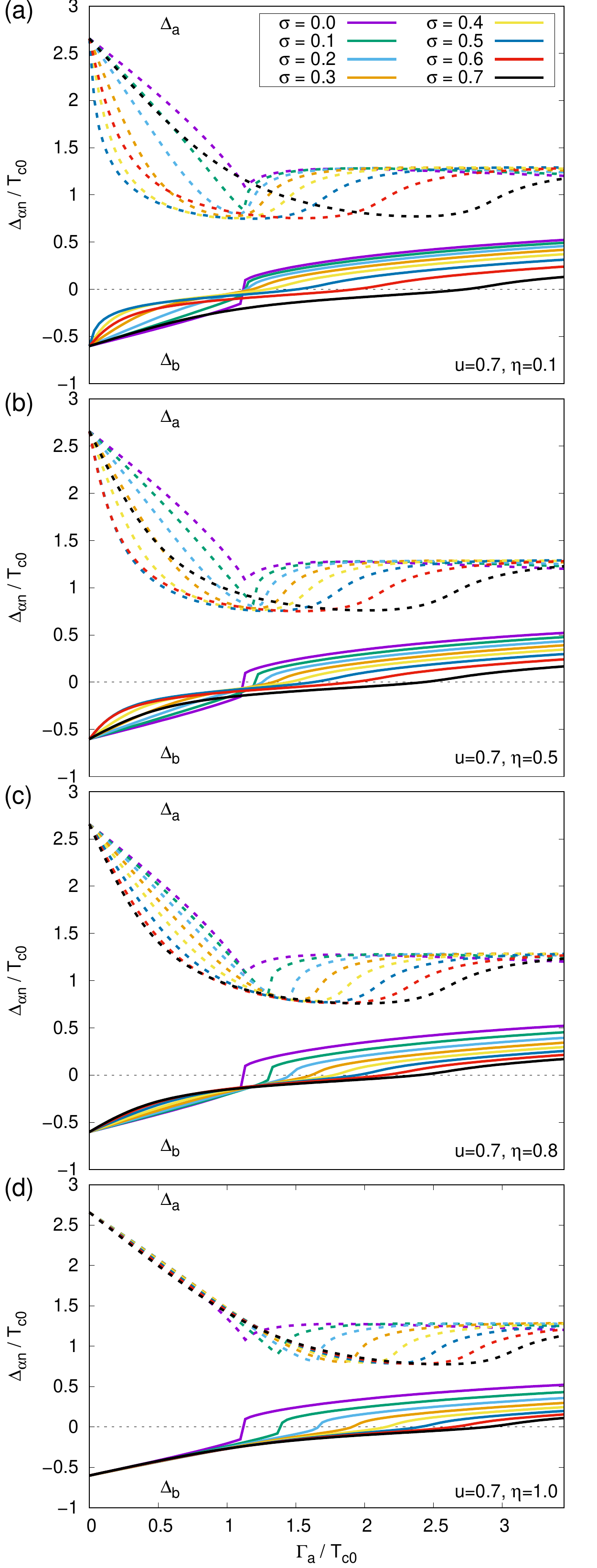}
 \caption{Matsubara gaps $\Delta_{\alpha n=1}$ vs. scattering rate $\GammaN_a$ in the $s_\pm$ state
 for various values of $\etaN$, i.e., $v=0.1u$ (a), $v=u/2$ (b), $v=0.8u$ (c), and uniform impurity potential $v=u$ (d).
 \label{fig:spmDeltaEtaAll}}
\end{figure}
\begin{figure}
 \centering
 \includegraphics[width=0.48\textwidth]{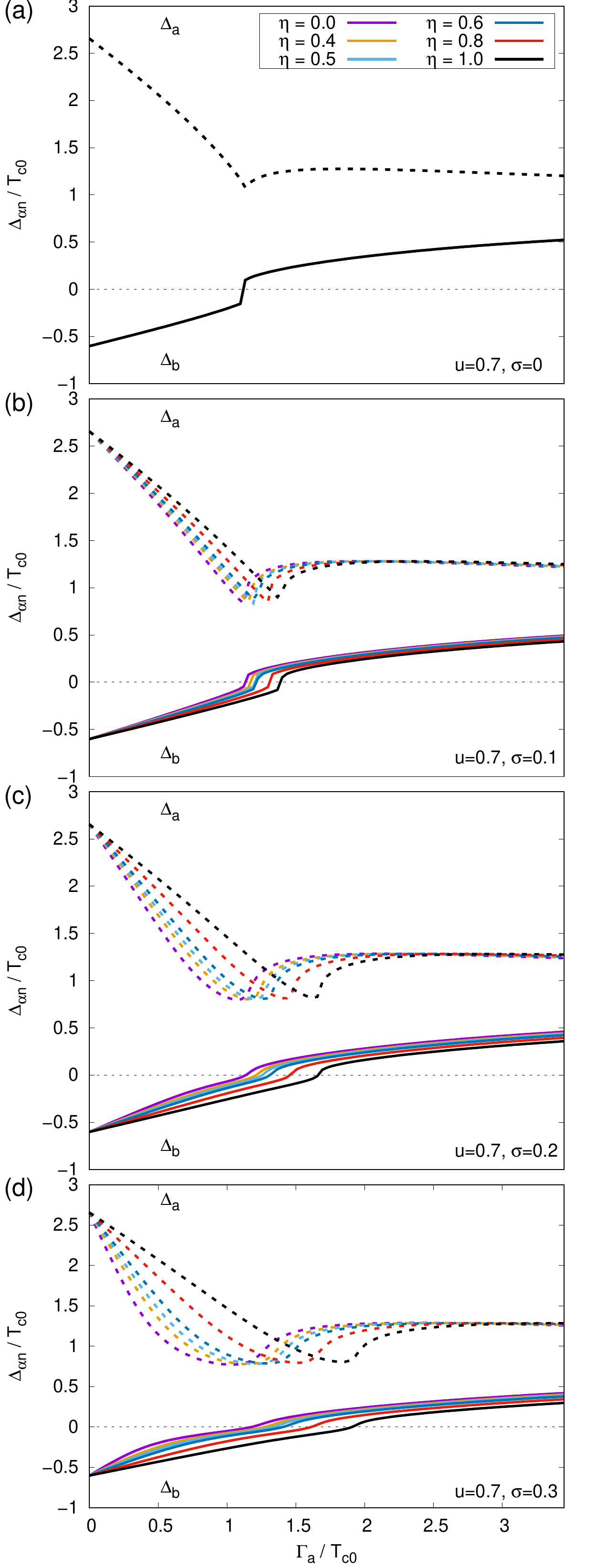}
 \caption{Matsubara gaps $\Delta_{\alpha n=1}$ vs. scattering rate $\GammaN_a$ in the $s_\pm$ state
 with varying $\etaN$ for $\sigmaN = 0$ (a), $\sigmaN = 0.1$ (b), $\sigmaN = 0.2$ (c), and $\sigmaN = 0.3$ (d). In panel (a), all curves for different $\etaN$ overlap.
 \label{fig:spmDeltaEtaSigma}}
\end{figure}

Here we choose the relation between densities of states as $N_b / N_a = 2$ and the following coupling constants: $(\lambda_{aa},\lambda_{ab},\lambda_{ba},\lambda_{bb}) = (3,-0.2,-0.1,0.5)$. This gives the $s_\pm$ state with the superconducting critical temperature $T_{c0}$ of 40~K
in the clean limit~\cite{Popovich2010,Charnukha2011,EfremovKorshunov2011} and the positive averaged coupling constant, $\la \lambda \ra > 0$, where $\la \lambda \ra \equiv (\lambda_{aa} + \lambda_{ab}) N_a / N + (\lambda_{ba} + \lambda_{bb}) N_b / N$~\cite{EfremovKorshunov2011}.

Since we are interested in behavior of gaps across the $s_\pm \to s_\pp$ transition, in Figure~\ref{fig:spmDeltaEta0} we plot $\Delta_{\alpha n}$ for the first Matsubara frequency $n=1$ as functions of $\GammaN_a$ at $T = 0.03 T_{c0}$. Following the $b$-band behavior, we observe the transition for $\GammaN_a \gtrsim 1.1 T_{c0}$. While for large values of $\etaN$ the gaps changes smoothly across zero, for $\sigmaN < 0.12$ we notice a jump in the smaller gap, $\Delta_{b n}$, when it crosses zero. It happens even in the Born limit, $\sigmaN = 0$. At the same time, critical temperature is always a smooth function of $\GammaN_a$, see the $T_c$ plot in Figure~\ref{fig:spmTcEta0}. Critical temperature seems to `do not care' about the jump occurring in the behavior of the smaller gap. To understand why this happens, we studied the temperature evolution of gaps. Results for $\sigmaN = 0$ and $\etaN = 0$ are shown in Figure~\ref{fig:spmDeltaT}. Apparently, with increasing temperature, the steep behavior of $\Delta_{b n}$ changes to the smooth dependence on the scattering rate. This happens at $T \sim 0.1 T_{c0}$ and, naturally, at higher temperatures, including $T_c$, the system shows smooth behavior. We have checked that the temperature dependence of gaps shown in Figure~\ref{fig:spmDeltaT} stays the same for $\etaN = 0.5$ and $\etaN = 1$.

It is known that the strongest $T_c$ suppression takes place in the Born limit with $\etaN = 0$, while in the opposite limit of pure intraband scattering with $u = 0$ ($\etaN \to \infty$), pairbreaking is absent because $\tilde\Gamma_{ab} \to 0$~\cite{KorshunovUFN2016}. Similar situation is also characteristic for the unitary limit with $\sigmaN = 1$, see equation~(\ref{eq.Gamma_ab}).

To demonstrate, how the transition evolves with the increasing intraband part of the impurity potential, $v$, in Figures~\ref{fig:spmDeltaEtaAll} and~\ref{fig:spmDeltaEtaSigma} we show results for different values of $\etaN$ at $T = 0.03 T_{c0}$. Thus, in Figure~\ref{fig:spmDeltaEtaAll}, we go from almost pure interband to uniform scattering. Apparently, the critical $\GammaN_a$ at which the transition takes place increases with increasing $v$ for $\sigma > 0$. In the Born limit, we observe the jump in the gaps for all $\etaN$'s at exactly the same critical $\GammaN_a$, see Figure~\ref{fig:spmDeltaEtaSigma}(a).

\section{Conclusions}

Here we studied the details of the $s_\pm \to s_\pp$ transition in the two-band model for the nonmagnetic impurity scattering in iron-based superconductors. We show that the gaps changes smoothly across the transition for all values of the cross-section parameter $\sigmaN$ and intra- to interband impurity potentials ratio $\etaN$, except for the case of a weak scattering with small values of $\sigmaN$. In the latter case, the smaller gap changes steeply at the transition point. For the larger scattering rate $\GammaN_a$, the smaller gap evolves smoothly.  The behavior changes around $\sigmaN = 0.11$.
With increasing temperature, the behavior of the gaps changes from the steep to the smooth around $T \sim 0.1 T_{c0}$ for all values of $\sigma$ and $\eta$. That is the reason why the critical temperature is always a smooth function of the scattering rate $\GammaN_a$ and doesn't care for steep changes in the behavior of the gaps.

\ack

We are grateful to A.V. Chubukov and P.D. Grigor'ev for useful discussions. This work was supported in part by the Russian Foundation for Basic Research (grant 16-02-00098) and Government Support of the Leading Scientific Schools of the Russian Federation (NSh-7559.2016.2). MMK and VAS acknowledge support of ``BASIS'' Foundation for Development of Theoretical Physics.

\section*{References}

\bibliographystyle{iopart-num}
\bibliography{mmkbibl3}

\end{document}